\documentclass[noshowkeys,tightenlines,nofootinbib, twocolumn]{revtex4}
\usepackage{epsfig}
\newif\ifpdf
\ifx\pdfoutput\undefined
\pdffalse % we are not running PDFLaTeX
\else
\pdfoutput=1 % we are running PDFLaTeX
\pdftrue
\fi

\def\Dsl{\hbox{/\kern-.6000em D}} %roman D

\def\dsl{\,\raise.15ex\hbox{/}\mkern-13.5mu D}

\def\ltap{\ \raise.3ex\hbox{$<$\kern-.75em\lower1ex\hbox{$\sim$}}\ }
\def\gtap{\ \raise.3ex\hbox{$>$\kern-.75em\lower1ex\hbox{$\sim$}}\ }
\def\OMIT#1{}

\def\OMIT#1{}

\newcommand{\nn}{\nonumber}

\newcommand{\bea}{\begin{eqnarray}}
\newcommand{\eea}{\end{eqnarray}}

\begin{document}

%%%%%%%%%%%%%%%%%%%%%%%%%%%%%%%%%%%%%%%%%%
%Some more stuff to get graphics to work
\ifpdf
\DeclareGraphicsExtensions{.pdf, .jpg}
\newcommand{\picspace}{\vspace{-2.5in}}
\newcommand{\picspacehalf}{\vspace{-1.75in}}
\else
\DeclareGraphicsExtensions{.eps, .jpg,.ps}
\newcommand{\picspace}{\vspace{0in}}
\newcommand{\picspacehalf}{\vspace{0in}}
\fi
%%%%%%%%%%%%%%%%%%%%%%%%%%%%%%%%%%%%%%%%%%

%%%%%%%%%%%%%%%%%%%%%%%%%%%%%%%%%%%%%%%%%%
%Define Title, Author, Address, Preprint#

%\preprint{\vbox{ \hbox{CMU-HEP-03-06}   \hbox{FERMILAB-Pub-03/069-T} }}

\title{ Implications of Heavy Quark-Diquark Symmetry for Excited Doubly Heavy Baryons and Tetraquarks}

\author{Thomas Mehen\footnote{Electronic address: mehen@phy.duke.edu}}
\affiliation{Department of Physics, Duke University, Durham,  NC 27708\vspace{0.2cm}} 

\date{\today\\ \vspace{1cm} }

%%%%%%%%%%%%%%%%%%%%%%%%%%%%%%%%%%%%%%%%%%

%%%%%%%%%%%%%%%%%%%%%%%%%%%%%%%%%%%%%%%%%%
%Create the title page

\begin{abstract}

We give heavy quark-diquark symmetry predictions for doubly heavy baryons and tetraquarks
in light of the recent discovery of the $\Xi_{cc}^{++}$ by LHCb. For five excited doubly charm baryons that are predicted to lie
below the $\Lambda_c D$ threshold, we give predictions for their electromagnetic and strong decays using a previously developed chiral Lagrangian with heavy quark-diquark symmetry. Based on the mass of the $\Xi_{cc}^{++}$,  the existence of a doubly heavy bottom $I=0$  tetraquark that is stable
to strong and electromagnetic decays has been predicted. If the mass of this state is below 10405 MeV, as predicted in some models, we argue 
using heavy quark-diquark symmetry that the $J^P=1^+$ $I=1$ doubly bottom tetraquark state will lie just below the open bottom threshold 
and likely be a narrow state as well. In this scenario, we compute strong decay width for this state using a new Lagrangian for doubly heavy tetraquarks which  is related 
to the singly heavy baryon Lagrangian by heavy quark-diquark symmetry.

\end{abstract}
\maketitle

%%%%%%%%%%%%%%%%%%%%%%%%%%%%%%%%%%%%%%%%%%
%\tighten
%\newpage
%%%%%%%%%%%%%%%%%%%%%%%%%%%%%%%%%%%%%%%%%%
%Main body of the paper

The LHCb experiment has recently observed  the doubly charm 
state, $\Xi_{cc}^{++}$\cite{Aaij:2017ueg}. While the SELEX collaboration~\cite{Mattson:2002vu, Moinester:2002uw,Ocherashvili:2004hi}
reported observations of doubly charm baryons years ago, these were not seen in other experiments and the
isospin violation implied by the recent LHCb measurement, 103 $\pm$ 2 MeV, seems implausibly large, casting doubt on the validity of these observations.
Important in confirming the nature of the LHCb discovery is observation of other excited doubly charm baryons. Quark models predict several excited 
states that lie below the open charm $\Lambda_c D$ threshold~\cite{Ebert:2002ig,Kiselev:2002iy,Kiselev:2001fw,Gershtein:2000nx,Gershtein:1998un,Gershtein:1998sx}. In this paper, we will study the five lowest doubly charm excitations and calculate their strong and electromagnetic decay widths using a chiral Lagrangian that exploits heavy quark-diquark symmetry first developed in Ref.~\cite{Hu:2005gf}. 

An interesting theoretical development that ensued after the LHCb discovery is the prediction of a
stable  doubly bottom $I=0$ tetraquark using a quark model~\cite{Karliner:2017qjm},
a mass formula based on heavy quark symmetry \cite{Eichten:2017ffp}, and lattice QCD calculations~\cite{Francis:2016hui,Bicudo:2016ooe}. An alternative argument for stability of doubly heavy tetraquarks is given in  Ref.~\cite{Czarnecki:2017vco}. An important point of this paper is to observe that 
if the mass of this tetraquark is less than 10405 MeV (as predicted in Ref.~\cite{Karliner:2017qjm}), then the lowest lying $J^P=1^+$ $I=1$ double bottom 
tetraquark is also likely to lie below the open bottom threshold. This state will decay strongly to the 
$I=0$ ground state by pion emission. In this paper, the mass and width of this state are estimated using a Lagrangian that uses heavy quark-diquark symmetry to relate doubly heavy tetraquarks to singly heavy baryons. 

In recent years the field of hadron spectroscopy has grown substantially, beginning with the discovery of the $X(3872)$ by the Belle collaboration in 2003~\cite{Choi:2003ue}. Since then numerous experiments   have observed  bottomonium and charmonium states, the so-called XYZ mesons, that do not fit into the conventional quark-antiquark potential model of quarkonium, for reviews see Refs.~\cite{Brambilla:2010cs,Esposito:2016noz,Guo:2017jvc,Ali:2017jda,Olsen:2017bmm}. Some of these states such as $Z_c^+(4430)$~\cite{Choi:2007wga,Aaij:2014jqa}  are charged and hence must exotic mesons. These could either be tetraquarks or molecular bound states of heavy-antiheavy mesons. Pentaquarks  have also been observed as resonances in $J/\psi p$ by the LHCb collaboration~\cite{Aaij:2015tga}.  Doubly charm baryons, while not exotic, are novel since no baryons with two heavy quarks have been definitively observed until Ref.~\cite{Aaij:2017ueg}. Some authors \cite{Esposito:2016noz} have posited a diquark-diquark bound state picture for many of the XYZ mesons. In a doubly heavy baryon, the heavy quark pair is expected to form a
compact diquark, so these baryons can shed some light on the dynamics of diquarks in quantum chromodynamics.

The heavy diquark in a doubly heavy hadron is a static source of color in the $\bf \bar{3}$ representation that looks the same to the light degrees of freedom as a singly heavy antiquark, up to corrections that are suppressed by the heavy quark mass.
Thus, heavy quark-diquark symmetry relates singly heavy mesons to doubly heavy baryons, and singly heavy baryons
to doubly heavy tetraquarks~\cite{Savage:1990di,Brambilla:2005yk,Fleming:2005pd,Cohen:2006jg,Flynn:2007qt,Brodsky:2011zs,Eakins:2012fq,Sun:2014aya,Wei:2015gsa,Ma:2015lba,Ma:2015cfa,Sun:2016wzh}. In Ref.~\cite{Hu:2005gf} a chiral Lagrangian for the ground state and excited 
states of doubly charm baryons was constructed. The ground state doubly charm baryon, which we identify with
the LHCb state, is denoted $\Xi_{cc}(3621)$. Heavy quark diquark symmetry predicts a spin partner, $\Xi^*_{cc}$, with mass splitting 
\bea\label{hyperfine}
m_{\Xi^*}-m_\Xi = \frac{3}{4}(m_{D^*}-m_D) \, ,
\eea
where $m_{D^*}$ and $m_D$ are the ground state charm vector and pseudoscalar meson masses, respectively.
Using this expression we expect an excited $\Xi^*_{cc}$ with a mass of 3727 MeV.  Lattice calculations of the doubly charm hyperfine splitting are consistent with this expectation within uncertainties~\cite{Lewis:2001iz,Mathur:2002ce,Flynn:2003vz,Chiu:2005zc,Na:2007pv,Na:2008hz,Liu:2009jc,Brown:2014ena}.
We assume isospin symmetry so the $\Xi_{cc}^{*++}$ and $\Xi_{cc}^{*+}$  are degenerate.
These states decay electromagnetically to the $\Xi_{cc}(3621)$ with widths $\Gamma[\Xi_{cc}^{*++}] = 5.1$ keV and $\Gamma[\Xi_{cc}^{*+}] = 7.6$ keV, where we use
the calculations of Ref.~\cite{Hu:2005gf} including non-analytic chiral loop corrections. There is an uncertainty of $\sim 40\%$ in these predictions coming from corrections due to heavy quark spin symmetry breaking.

The lowest lying excitations above the ground state doublet are internal excitations of the $cc$ diquark rather than the light degrees of freedom in the baryon. In the quark model, the lowest excitations  of the light degrees of freedom correspond to states in which the light quark has one unit of angular momentum.  In $D$ mesons these excitation are $\sim450$ MeV above the ground state
doublet. The corresponding excitations in the doubly charm sector will be well above the $\Lambda_c D$ threshold.
Because the potential between the $c$ quarks in the diquark is 1/2 as strong as the potential between the charm and anticharm quark in charmonium, the internal excitations of the $cc$ diquark are expected to be approximately 1/2 the size of the corresponding charmonium excitations. The two lowest lying excitations are states with an $S=0$ $cc$ diquark in a $P$-wave and the first radially excited diquark. Quark models anticipate these excitation energies to be 225 MeV and 300 MeV, 
respectively~\cite{Ebert:2002ig,Kiselev:2002iy,Kiselev:2001fw,Gershtein:2000nx,Gershtein:1998un,Gershtein:1998sx}, and we will assume these excitation energies in what follows.
We could arrive at very similar predictions by noting that $(m_{h_c}-m_{J/\psi})/2 = 214$ MeV and $(m_{\psi(2S)}-m_{J/\psi})/2 = 295$ MeV. We also assume the hyperfine 
splittings of these multiplets are the same as the hyperfine splittings in the ground state doublet. Since the light degrees of freedom are the same as the ground state, the formula in Eq.~(\ref{hyperfine}) should hold for these doublets as well. Quark model calculations also give the same hyperfine splittings for all three doublets~\cite{Ebert:2002ig,Kiselev:2002iy,Kiselev:2001fw,Gershtein:2000nx,Gershtein:1998un,Gershtein:1998sx}. We denote the baryons containing $P$-wave excitation of the diquark as $\Xi_{cc}^{\cal P}$ and $\Xi_{cc}^{\cal P*}$, these states have $J=1/2$ and $J=3/2$, respectively. We denote the baryon with a radial excited diquark  $\Xi_{cc}^{\prime}$ and $\Xi_{cc}^{\prime *}$, these states also have $J=1/2$ and $J=3/2$, respectively.  The assumed masses of the five excited doubly charm baryons are: $m_{\Xi^*_{cc}}=3727$ MeV, $m_{\Xi^{\cal P}_{cc}}=3846$ MeV, $m_{\Xi^{\cal P*}_{cc}}=3952$ MeV, $m_{\Xi^{\prime}_{cc}}=3921$ MeV, and $m_{\Xi^{\prime*}_{cc}}=4027$ MeV.
Note that for the assumed excitation energies all of these states are below the open $\Lambda_c^+ D$ threshold of $\approx 4150$ MeV. Therefore the excited  doubly charm baryons must decay to the $\Xi_{cc}(3621)$ or $\Xi_{cc}^*$ via pion emission, and can be narrow. 

The widths for $\Xi_{cc}^{\cal P}$ and $\Xi_{cc}^{\cal P*}$ were calculated in Ref.~\cite{Hu:2005gf}:
\bea
\Gamma[\Xi_{cc}^{\cal P *} \to \Xi_{cc}^* \, \pi] &=&  \lambda_{3/2}^2 \, 112 \,{\rm MeV} \nn \\
\Gamma[\Xi_{cc}^{\cal P} \to \Xi_{cc} \, \pi] &=&  \lambda_{1/2}^2 \,  111 \,{\rm MeV} \, ,  
\eea
where $\lambda_{1/2}$ and $\lambda_{3/2}$ are coupling constants defined in Ref.~\cite{Hu:2005gf}.\footnote{In the Appendix we give an argument that $\lambda_{1/2}=\lambda_{3/2}$ in the heavy quark limit, but this could receive substantial corrections in the charm sector.}
Note that these decays violate heavy quark spin symmetry as the 
spin of the diquark changes in this transition. Thus $\lambda_{1/2},\lambda_{3/2} \sim \Lambda_{\rm QCD}/m_c \ll 1$. For $1/2<\lambda_{1/2},\lambda_{3/2} <1/3$
 the widths  are in the range 12 -28 MeV. For the radial excitations $\Xi_{cc}^{\prime}$ and $\Xi_{cc}^{\prime}$, these can decay to either of the ground state baryons. Using the assumed excitation energies, application of the formulae from 
  Ref.~\cite{Hu:2005gf} yields
 \bea 
\Gamma[\Xi_{cc}^{\prime}] =  {\tilde g}^2 \, 52 \, {\rm MeV} \qquad&& \Gamma[\Xi_{cc}^{\prime *}] = {\tilde g}^2 \,391 \, {\rm
MeV} \nn \\ 
&&\nn \\  \frac{\Gamma[\Xi_{cc}^{\prime *} \to \Xi_{cc}^*  \, \pi]}{\Gamma[\Xi_{cc}^{\prime *} \to \Xi_{cc} \, \pi]} = 0.46  \qquad &&
\frac{\Gamma[\Xi_{cc}^{\prime} \to \Xi_{cc}^* \,  \pi]}{\Gamma[\Xi_{cc}^{\prime} \to \Xi_{cc} \,  \pi]} = 1.2  , \,
\eea 
 where $\tilde g$ is an unknown coupling constant of order unity. $\Xi_{cc}^{\prime *}$ decays to the ground state $\approx 2/3$ of the time
 while $\Xi_{cc}^{\prime}$ decays to either of the ground state with roughly equal probability. Note that these ratios are predicted by heavy quark symmetry, including corrections for the phase space factors on the decay. $\Xi_{cc}^{\prime*}$ is expected to be broad while $\Xi_{cc}^\prime$ is expected 
 to be significantly narrower.
  
So far only doubly heavy baryons with light $u,d$ quarks have been considered. It is straightforward to extend the Lagrangian
in Ref.~\cite{Hu:2005gf}  to include doubly heavy baryons with strangeness. A mass term 
 linear in quark masses would give the leading $SU(3)$ breaking to the masses. An obvious prediction of heavy quark-diquark symmetry (in the isospin limit) is
 \bea
 m_{\Omega_{cc}} -m_{\Xi_{cc}^{+,++}} = m_{D_s} -m_{D^{0,+}} = 101 \, {\rm MeV} \, ,
 \eea
where $\Omega_{cc}$ is the ground state doubly charm, strange baryon.  This prediction is consistent with the lattice calculation of $m_{\Omega_{cc}} -m_{\Xi_{cc}^{+,++}} = 98\pm9\pm22\pm13$  in Ref.~\cite{Liu:2009jc}. Observation of the $\Omega_{cc}$ with a mass close to 3722 MeV would provide further evidence for heavy quark-diquark symmetry. 
The expected low lying excitation spectrum in the strange sector is nearly identical to that in the nonstrange sector.
All five excited doubly charm strange baryons would lie below threshold for decay via kaon emission. The strong decays of 
the baryons with $P$-wave $cc$ diquarks  to the ground state via  isospin violating $\pi^0$ emission were calculated in Ref.~\cite{Hu:2005gf} and their widths are expected to be a few keV. Baryon with radially excited $cc$ diquarks have similar strong decays and are expected to be narrow.

 For studying doubly bottom tetraquarks we will develop a chiral Lagrangian by applying heavy quark-diquark symmetry to the chiral Lagrangian for singly heavy 
baryons including nonstrange baryons only  \cite{Cho:1992cf,Cho:1992gg}. The relevant terms in the chiral Lagrangian for singly heavy baryons can be written as 
\bea\label{baryonLag}
{\cal L}_B &=& \bar{\Sigma}^{\alpha,i} iD_0 \Sigma^{\alpha,i} + \bar{\Lambda} i D_0 \Lambda -(m_{\bar \Sigma} -m_\Lambda)  \bar{\Sigma}^{\alpha,i} \Sigma^{\alpha,i}  
 \nn \\
&&+i \Delta \bar{\Sigma}^{\alpha,i}\epsilon_{ijk} \sigma^k \Sigma^{\alpha,j} -g_3 (\bar{\Lambda} A^{\alpha,i} \Sigma^{\alpha,i} + h.c.) \, . 
\eea
Here we are working in the rest frame of the heavy baryon so its four-velocity is $v^\mu=(1,\vec{0})$. In this Lagrangian a bar over a field denotes hermitian conjugate, $D_0$ is a chirally covariant derivative, the axial vector field is  $A^{\alpha,i} = -\partial^i \pi^\alpha/f + ...$, where the superscript $i$ is a vector index, $\alpha$ is an SU(2) vector index,  $f = 131$ MeV is the pion decay constant, $\pi^\alpha$ is the pion field, $\Lambda$ is a two-component spinor field for the $SU(2)$ singlet $\Lambda$ baryon, and $\Sigma^{\alpha,i}$ is a vector-spinor isovector field for the $\Sigma$ baryons. This field can be further decomposed as 
\bea
\Sigma^{\alpha,i} = \Sigma^{*\alpha,i} + \frac{\sigma^i}{\sqrt{3}}\Sigma^\alpha \, ,
\eea
where $\Sigma^\alpha$ are the $J=\frac{1}{2}$ $\Sigma$ baryons and $\Sigma^{*\alpha,i}$ are the $J=\frac{3}{2}$ $\Sigma^*$ baryons and obey the constraint $\sigma^i \Sigma^{*\alpha,i}=0$.
The first two terms in Eq.~(\ref{baryonLag}) are the kinetic terms for the baryons, the third term gives the $\Sigma-\Lambda$ mass splitting, the fourth term gives the hyerfine mass splitting
between the $\Sigma^*$ and $\Sigma$ and the last term gives the coupling responsible for the decay $\Sigma \to \Lambda \pi$. 

The corresponding Lagrangian for doubly heavy tetraquarks is given by making the substitutions
\bea
\Lambda &\to& T_\Lambda^j \nn \\
\Sigma^{\alpha,i} &\to& T_\Sigma^{\alpha,ji} \, ,
\eea
where $T_\Lambda^j$ and $T_\Sigma^{\alpha,ji}$ are fields for the doubly heavy tetraquarks and $j$ is the spin index associated with the spin-1 diquark. Under rotations 
and heavy quark spin transformations these fields transform as 
\bea
T_\Lambda^j \to R_Q^{jk} T_\Lambda^k && \qquad T_\Sigma^{\alpha,ji} \to R_Q^{jk} T_\Sigma^{\alpha,ki} \nn \\
T_\Lambda^j \to R^{jk} T_\Lambda^k  &&\qquad T_\Sigma^{\alpha,ji} \to R^{jk} R^{il} T_\Sigma^{\alpha,kl} \, ,
\eea 
where $R_Q$ is a heavy quark spin rotation and $R$ is a rotation matrix. The field $T_\Sigma^{\alpha, ji}$ is reducible:
\bea
T_\Sigma^{\alpha, ji} = T_{\Sigma^{**}}^{\alpha, ji} + \frac{1}{\sqrt{2}}\epsilon^{jik}T_{\Sigma^*}^k +\frac{\delta^{ij}}{\sqrt{3}} T_\Sigma \, ,
\eea
where $T_{\Sigma^{**}}, T_{\Sigma^{*}}$, and $T_\Sigma$ are the spin-2, spin-1, and spin-0 fields, respectively. In the term in the Lagrangian responsible for the 
hyperfine splitting we also make the substitution $\sigma^i \to -i \epsilon_{ilm}$, as explained in Refs.~\cite{Fleming:2005pd,Hu:2005gf}. Here the indices $lm$ are contracted with the heavy quark spin indices in the fields, so this term in the Lagrangian for doubly heavy tetraquarks is
\bea
{\cal L}_{hf}  &=& \Delta \bar{T}_\Sigma^{\alpha,li} \epsilon_{ijk}\epsilon_{klm} T_\Sigma^{\alpha,mj} \nn \\
&=& -\Delta \bar{T}_{\Sigma^{**}}^{\alpha,ij} T_{\Sigma^{**}}^{\alpha,ij} +\Delta \bar{T}_{\Sigma^*}^k T_{\Sigma^*}^k+2 \Delta \bar{T}_\Sigma T_\Sigma \, .
\eea
We find 
\bea\label{mass}
 m_{T_\Sigma}-m_{T_\Lambda} &=& m_\Sigma -m_\Lambda  \nn \\
 m_{T_{\Sigma^{**}}}-m_{T_{\Sigma^{*}}} &=& \frac{2}{3}( m_{\Sigma^*} -m_\Sigma) \nn \\ 
m_{T_{\Sigma^{*}}}-m_{T_\Sigma} &=& \frac{1}{3}( m_{\Sigma^*} -m_\Sigma) \, .
\eea
From these formulae and the known masses of the $\Sigma_{b,c}, \Sigma_{b,c}^*$ and $\Lambda_{b,c}$, we conclude 
that for doubly bottom tetraquarks,
\bea
 m_{T_{\Sigma_{bb}}} &=&m_{T_{\Lambda_{bb}}} +193 \, {\rm MeV}  \nn \\
 m_{T_{\Sigma^*_{bb}}} &=&m_{T_{\Lambda_{bb}}} +200 \, {\rm MeV}  \nn \\
 m_{T_{\Sigma^{**}_{bb}}} &=&m_{T_{\Lambda_{bb}}} +214 \, {\rm MeV} \, ,
\eea
and for doubly charm tetraquarks
\bea
 m_{T_{\Sigma_{cc}}} &=&m_{T_{\Lambda_{cc}}} +168 \, {\rm MeV}  \nn \\
  m_{T_{\Sigma^*_{cc}}} &=&m_{T_{\Lambda_{cc}}} + 189 \, {\rm MeV}  \nn \\
    m_{T_{\Sigma^{**}_{cc}}} &=&m_{T_{\Lambda_{cc}}} + 232 \, {\rm MeV} \, .
\eea
For the doubly charm baryons, $m_{T_{\Lambda_{cc}}}$ is expected to be above the open charm threshold \cite{Karliner:2017qjm,Eichten:2017ffp} 
so the $T_{\Sigma_{cc}}$  will be hundreds of MeV above threshold. These will be broad states and the strong decay will be dominated by decays 
to open charm. For $T_{\Lambda_{bb}}$, Ref.~\cite{Karliner:2017qjm} predicts $m_{T_{\Lambda_{bb}}} = 10389\pm12$ MeV, which yields $m_{T_{\Sigma^*_{bb}}} =10589$ MeV for the central value. This is below the $m_B+m_{B^*}$ threshold which is the relevant threshold since $T_{\Sigma^*_{bb}}$ is spin 1. This state will be narrow because open bottom decay channels are closed and will decay to $T_{\Lambda_b} \pi$. This width is calculated below. The $T_{\Sigma^{**}_{bb}}$ and $T_{\Sigma_{bb}}$ can decay to two pseudoscalars and lie well above this threshold. These states will be broad and their strong widths dominated by decays to open bottom.

Let us comment the accuracy of the predictions in Eqs.~(\ref{mass}). These formulae imply the following relationship between the spin averaged mass of the $\Sigma$ and $T_\Sigma$
multiplets and the  $\Lambda$ and $T_\Lambda$ states:
\bea\label{spinave}
m_{T_{\bar \Sigma}}-m_{T_\Lambda} &=& m_{\bar\Sigma} -m_\Lambda 
\eea
where  $m_{T_{\bar \Sigma}}= (m_{T_{\Sigma}}+3 m_{T_{\Sigma^*}}+5 m_{T_{\Sigma^{**}}})/9$ and $m_{\bar\Sigma} =(2 m_{\Sigma^*}+m_\Sigma)/3$. This relation is essentially 
saying that replacing an $I=0$ $S=0$ diquark with a $I=1$ $S=1$ will increase the energy of the baryon by an amount that is independent off the heavy quark. There is plenty of evidence 
to support this from the charm, bottom, and even strange sectors, since
\bea\label{massdiff}
m_{\bar \Sigma_b} - m_{\Lambda_b} &=& 207 \, {\rm MeV} \nn \\
 m_{\bar \Sigma_c} - m_{\Lambda_c} &=& 210 \, {\rm MeV}  \nn \\
 m_{\bar \Sigma} - m_\Lambda &=& 205 \, {\rm MeV}  \, ,
 \eea
where we have used isospin averaged masses in computing Eqs.~(\ref{massdiff}). We see that the excitation energy of $I=1$ $S=1$ diquark is consistently $\approx 207$ MeV
within a few MeV, so we feel confident  Eq.~(\ref{spinave}) will continue to hold in the doubly charm and doubly bottom sector. The second aspect of our predictions is  the relationship between hyperfine splittings in the singly heavy baryon and doubly heavy meson sector in Eqs.(\ref{mass}). These have not been tested by experimental data, and could receive large corrections, particularly in the double charm sector. However, these give small shifts to the masses relative to that in Eq.~(\ref{spinave}). In the doubly bottom sector, the  hyperfine 
splittings shift the $T_{\Sigma^{**}_{bb}}$ to 7 MeV above the spin averaged mass, $T_{\Sigma^{*}_{bb}}$ is 7 MeV below, and $T_{\Sigma_{bb}}$ 14 MeV below. In the doubly charm sector these splittings are $\pm 21$ MeV and $-42$ MeV. So even if heavy quark-diquark symmetry predictions for hyperfine splitting receive 50\% corrections, this will only change masses in the doubly bottom sector by 4-7 MeV, and in the doubly charm sector by 10-21 MeV. Therefore, our conclusions about the spectrum of these excitations should be robust.

We will estimate the strong width for $T_{\Sigma^*_{bb}}$ in the scenario where $m_{T_{\Lambda_{bb}}} = 10389$ MeV and $m_{T_{\Sigma^*_{bb}}} = 10589$ MeV.
The decay rates for $T_{\Sigma^{(*,**)}} \to T_\Lambda \pi$  from the chiral Lagrangian are:
\bea
\Gamma[T_{\Sigma^{**}} \to T_\Lambda \pi] &=& \Gamma[T_{\Sigma^{*}} \to T_\Lambda \pi] 
 = \Gamma[T_\Sigma \to T_\Lambda \pi] \nn \\
 &=& \frac{ g_3^2}{6 \pi f^2} \frac{m_{T_\Lambda}}{m_{T_{\Sigma}}} p_\pi^3 \,.
 \eea
The prediction for the decay rate is the same as the rate for $\Gamma[\Sigma^{(*)} \to \Lambda_c^+ \pi]$ found in Ref.~\cite{Cho:1992gg}.
This is the only strong decay  channel for the $\Sigma_c$ and $\Sigma_c^*$, their widths are  $\Gamma[\Sigma_c] = 1.86^{+0.10}_{-0.19}$ MeV
and $\Gamma[\Sigma^*_c] = 15.0^{+0.4}_{-0.5}$, respectively,  where we have averaged over the two isospin channels for which  there is a measurement. 
These widths are consistent with $g_3^2 = 0.93$ with a few percent error. For this value of $g_3$ the predicted widths for the $b$ sector are 
$\Gamma[\Sigma_b] = 6.3$ MeV and $\Gamma[\Sigma^*_b] = 11.3$ MeV, which  are consistent with measured widths but in this case the experimental uncertainties are much larger~\cite{Olive:2016xmw}.
In the scenario where $m_{T_{\Lambda_{bb}}} = 10389$ MeV, which yields $m_{T_{\Sigma^*_{bb}}} =10589$ MeV, using $g_3^2 = 0.93$,
we find 
\bea\label{rates}
\Gamma [ T_{\Sigma_{bb}^{**}} \to T_{\Lambda_{bb}} \pi ]  &=& 11.8 \, {\rm MeV} \nn \\
 \Gamma[T_{\Sigma_{bb}^{*}} \to T_{\Lambda_{bb}} \pi ]  &=& 8.1\, {\rm MeV} \nn \\  
 \Gamma[T_{\Sigma_{bb}} \to T_{\Lambda_{bb}} \pi ]  &=& 6.5\, {\rm MeV} \, ,
 \eea
 where the corrections to these estimates are $O(\Lambda_{\rm QCD}/m_c)$ since we are extracting the coupling from the single charm sector. Of course 
 the total widths for $T_{\Sigma_{bb}}$ and $T_{\Sigma^{**}_{bb}}$ will receive substantial corrections due to open bottom decay channels. The $T_{\Sigma^*_{bb}}$ is a narrow state with a width less than $10$ MeV.

To summarize, we have considered the five lowest lying excitations of the recently observed $\Xi_{cc}^{++}$, and its isospin partner, and predicted their strong and electromagnetic decay rates using a chiral Lagrangian with heavy quark-diquark symmetry first derived in Ref.~\cite{Hu:2005gf}. We then wrote down a chiral Lagrangian for heavy tetraquarks 
that uses heavy-quark diquark symmetry to relate properties of doubly heavy tetraquarks to the singly heavy $\Sigma_{c,b}$ and $\Lambda_{c,b}$ baryons. Most excited doubly heavy tetraquarks 
are above the open charm and bottom thresholds. If the $T_{\Lambda_{bb}}$ is less than 10405 MeV, then the $T_{\Sigma^*_{bb}}$ is predicted to be below the $BB^*$ threshold and 
 will decay to the $T_{\Lambda_{bb}}$ via single pion emission and have a width $<10$ MeV. It may be interesting to extend the calculations of this paper to tetraquarks with strangeness, as well as include corrections to isospin symmetry and/or $SU(3)$. 
 
 While this paper was in preparation, Ref.~\cite{Li:2017pxa} appeared which also computed electromagnetic decays of doubly heavy baryons using chiral perturbation theory,
 as well as Refs.~\cite{Lu:2017meb,Xiao:2017udy} which studied strong and radiative decays of doubly charm baryons using quark models. 

\begin{acknowledgments}

This research is supported in part by  Director, Office of Science, Office of Nuclear Physics,
of the U.S. Department of Energy under grant number
DE-FG02-05ER41368. 

 \end{acknowledgments}

\section{Appendix}

Ref.~\cite{Hu:2005gf} argued that the $\Xi_{cc}^{\cal P}$ and  $\Xi_{cc}^{\cal P*}$ could not be placed in a heavy quark spin symmetry multiplet because they contain a $S=0$ $L=1$ diquark. The simplest Lagrangian for mediating $S$-wave decays to the ground state was written down and the couplings for these two decays were considered independent. However, it is clear that in the heavy quark limit  these states are degenerate and these decays should be related. If we generalize the notion of heavy quark spin symmetry to include orbital angular momentum of the heavy quarks, then the diquark is a vector under rotations generated by $\vec{J}_Q = \vec{S}_Q+\vec{L}_Q$, where $\vec{J}_Q$, $\vec{S}_Q$,
and $\vec{L}_Q$ are the heavy quark total, spin, and orbital angular momentum, respectively. The  $\Xi_{cc}^{\cal P}$ and  $\Xi_{cc}^{\cal P*}$ are now in a multiplet just like $\Xi_{cc}$ and  $\Xi_{cc}^*$.
A coupling that is invariant under rotations generated by $\vec{J}_Q$ (but not $\vec{S}_Q$ or $\vec{L}_Q$ separately) is
\bea\label{last}
{\cal L} &=& \lambda T_{a,i\beta}^\dagger  T^{\cal P}_{b,i\beta} A^0_{ba} \nn \\
&=& 2 \lambda\,  \Xi_a^\dagger \Xi_b^{\cal P} A^0_{ba} + 2\lambda \, \Xi_a^{*\dagger} \Xi^{\cal P*}_b A^0_{ba}  +{\rm h.c.}\, ,
\eea
where $T_{a,i\beta}$ is the superfield defined in Eq.~(13) of Ref.~\cite{Hu:2005gf}, $T^{\cal P}_{a,i\beta}$ is an analogous superfield with the odd parity baryons, and the 
second line of Eq.~(\ref{last}) is the same as the Lagrangian in Ref.~\cite{Hu:2005gf} with $\lambda_{1/2}=\lambda_{3/2}=\lambda$. Since the operator flips this spin of the 
heavy quark the coupling $\lambda$ is $O(\Lambda_{\rm QCD}/m_Q)$. Deviations from the prediction 
$\Gamma[\Xi_{cc}^{\cal P *} \to \Xi_{cc}^* \, \pi]/\Gamma[\Xi_{cc}^{\cal P} \to \Xi_{cc} \, \pi]=1$ will indicate the size of heavy quark symmetry breaking effects in excited doubly charm baryons.

%%%%%%%%%%%%%%%%%%%%%%%%%%%%%%%%%%%%%%%%%%a
%Bibliography
\bibliography{paper}

\end{document}